\documentclass{ws-ijmpcs}

\def\gsim{ \lower .75ex \hbox{$\sim$} \llap{\raise .27ex \hbox{$>$}} }
\def\lsim{ \lower .75ex\hbox{$\sim$} \llap{\raise .27ex \hbox{$<$}} }

\def\sc{Schwarzschild}
\def\beq{\begin{equation}}
\def\eeq{\end{equation}}

\def\sc{Schwarzschild}

\begin{document}

\markboth{Gabriele Ghisellini}
{Jetted AGNs}

%
\catchline{}{}{}{}{}
%

\title{JETTED ACTIVE GALACTIC NUCLEI}

\author{Gabriele Ghisellini}

\address{INAF -- Osservatorio Astronomico di Brera, Via Bianchi 46 I--23807  Merate Italy
\footnote{gabriele.ghisellini@brera.inaf.it}\\
}

\maketitle

\begin{history}
\received{}
\revised{}
\end{history}

\begin{abstract}
Most of the electromagnetic output of blazars (BL Lac objects and Flat 
Spectrum Radio Quasars) comes out in the $\gamma$--ray band, making the 
Large Area Telescope [0.1--100 GeV] onboard the {\it Fermi} satellite and the 
Cherenkov telescopes crucial for gather crucial data and thus to understand 
their physics. 
These data are complemented by the observations of the 
{\it Swift} satellite in the X--ray and optical--UV bands, and by ground 
based optical and radio telescopes. This rich coverage of the spectrum 
allows a robust modelling, from which important trends start to emerge. 
In powerful sources we see the contribution of the accretion disk
that, once modeled, give us the black hole mass and the accretion rate.
Even when not directly visible, the disk luminosity can be derived 
through the broad emission lines.
Therefore we start to know the jet power, the disk luminosity, and the
black hole mass, 3 crucial ingredients if we want to draw a general scenario.
At the start, jets are believed to be magnetically dominated. 
And yet, on the scale where they emit most of their luminosity, their power is 
already in the form of kinetic energy of particles.
Relativistic jets are formed for a very broad range of the disk luminosity,
from close to Eddington down to at least $10^{-4}$ Eddington.
Their power correlates with the accretion rate, and can be even more 
powerful than the accretion disk luminosity.

\end{abstract}

\ccode{PACS numbers: 98.54.Cm,  98.54.Aj, 98.70.Rz }

\section{Introduction}	
The jets of Flat Spectrum Radio Quasars (FSRQs) and BL Lac objects,
being relativistic and pointing at us, are the most powerful persistent
sources in the Universe.
After decades of research we have still to explain the main
mechanism launching, accelerating and collimating jets, but part of 
our partial lack of progress can be ascribed to our ignorance 
of basic physical parameters, such as the power of jets, their composition,
the particle acceleration mechanisms and the efficiency 
in producing the radiation we see.
In this respect, the {\it Compton Gamma Ray Observatory (CGRO)} with
its high energy instrument, EGRET, and the ground based Cherenkov telescopes were 
crucial to unveil at least the entire apparent (i.e. isotropic equivalent) emitted 
luminosity, that often peaks in the $\gamma$--ray band.
Now, with more advanced Cherenkov telescopes and especially with the
successor of {\it CGRO}, namely the {\it Fermi} satellite and its
Large Area Telescope (LAT, 20 times more sensitive than EGRET), we
entered in a new era (e.g. Ref. \refcite{abdo2009}, \refcite{abdo2010a}, \refcite{abdo2010b}).
Besides the improved capabilities of these instruments, we also
benefit from the flexibility of the {\it Swift} satellite, that makes possible 
to have routinely optical, near UV and X--ray data
really simultaneously with the high energy ones.

Furthermore, being visible up to large redshifts, blazars 
can be used as probes of the far Universe even if do not understand
completely their physics.
We can exploit the fact that the Universe is not transparent to high energy 
($\sim$TeV) $\gamma$--rays, because of the cosmic IR background photons
that can interact with $\gamma$--rays and absorb them, leaving 
some fingerprint of their presence.
And, more recently, blazars open the search for the ``cosmic magnetic field background".

Prospects are rather optimistic for blazar research,
and in the following I will try to outline some of the results 
already obtained.

\section{Low power BL Lacs}

At the low end of the blazar sequence\cite{fossati1998}\cdash\cite{donato2001}
we find low power, TeV emitting, BL Lac objects.
The TeV flux of PKS 2155--304\cite{aharonian2007} and Mkn 501\cite{albert2007}
showed, occasionally, ultrafast
variability, with a doubling time of the order of 3--5 minutes.
This, as discussed below, is a problem for our naive expectations,
and challenges all models.
Even more intriguing, and challenging, is the recent finding that not only 
lineless, low power BL Lacs vary fast, but even broad emitting line FSRQs do\cite{aleksic2011}.

TeV photons can be converted to $e^\pm$ pairs while traveling to Earth, by interacting
with the cosmic optical--IR background photons. 
The resulting absorption features have been used to test models of the optical--IR backgrounds, 
\cite{orr2011}\cdash\cite{primack2011}\cdash\cite{costamante2003}
but a more recent use of this process concerns the estimate of the cosmic magnetic field.

\subsection{Ultrafast TeV variability}

Theoretically, one expects that the minimum variability timescale in jetted AGNs
(and in Gamma Ray Bursts, too) is linked to the light crossing time of the \sc\
radius of the black hole, even if the emission region is very distant from the
central engine.
As an example, consider the ``internal shock scenario", popular in the Gamma Ray Burst
field, but applicable to blazars too (for which, indeed, it was originally invented\cite{rees1978}).
Assume that two shells are initially separated by $\Delta R$ and have a similar width.
Take $\Delta R$ of the order of the \sc\ radius $R_{\rm S}$.
Assume the front one moves with a bulk Lorentz factor $\Gamma$ and the 
back one moves with $2\Gamma$.
They will collide at a distance $R_{\rm coll}\sim \Gamma^2\Delta R  \sim \Gamma^2 R_{\rm S}$, 
and will interact for a distance equal to $R_{\rm coll}$.
For observers on the jet axis, the Doppler contraction will shorten the observed time
to $t_{\rm var} = R_{\rm coll}/(c \Gamma^2) \sim R_{\rm S}/c$.
A similar timescale is expected for conical jets producing TeV photons at a distance $\sim 10^3 R_{\rm S}$
from the central engine, from a region with a size equal to the cross sectional radius of the jet.
For a $10^9 M_\odot$ black hole, we expect $t_{\rm var} \sim R_{\rm S}/c\sim$3 hours.
We instead see hundreds of seconds, corresponding to a size $R \lsim c t_{\rm var} \Gamma
\sim 3\times 10^{13} t_{\rm var, 2}\Gamma_1$ cm.
We then require very compact regions moving very fast. 
One possibility is to have small reconnecting regions, producing anisotropic 
currents of relativistic electrons in the comoving frame of the jet.
We then have a ``jet in a jet" scenario\cite{giannios2009} (relativistic flows of particles in a frame 
that is also moving relativistically).
These electrons would inverse Compton scatter their own synchrotron radiation or any ambient photon
field, cool very rapidly, and produce fast variations. 
An alternative model is to invoke magneto--centrifugal acceleration of electrons along
rigid and rotating $B$--field lines\cite{ghisellini2010}. 
Electrons would reach their maximum speed close to the ``normal" dissipation region,
and could inverse Compton scatter very efficiently the synchrotron photons of that region. 
These electrons would move in a very collimated beam, and relativistic effect would be huge.
A strong requirement of this model is that the accretion disk must be radiatively
inefficient, since otherwise it would prevent the accelerating electrons to reach
the required energies.
In BL Lacs we do have inefficient disks, and so one would be tempted to associate 
ultrafast TeV variability to BL Lac objects only.

The recent detection\cite{aleksic2011} of ultrafast variability ($\sim$10 minutes) from a FSRQ 
source  by MAGIC (PKS 1222+216, $z=0.431$) came therefore as a surprise.
To let TeV photons survive, the emitting region should be away from the accretion disk
(that in this source is efficient, we see strong broad lines), and yet we need a
compact source.
In this case electrons accelerated by the magneto--centrifugal force
cannot reach energies high enough to produce TeV photons.
On the other hand, also the ``jet in a jet" reconnection model
may have problems, since it requires a magnetically dominated jet, 
while the synchrotron luminosity of this source is not 
dominant (see Ref. \refcite{tavecchio1222}).

\subsection{Cosmic magnetic fields}

The TeV photons produced by blazars, absorbed by the IR background,
produce $e^\pm$ pairs, that efficiently cool by Inverse Compton scattering
off the CMB photons.
Photons of energy $\epsilon=10$ TeV would produce pairs of 5 TeV each, 
corresponding to a $\gamma$--factor of $\sim 10^7$. 
Their reprocessed radiation will have $\epsilon\sim 100$ GeV.

\begin{figure} 
\vskip -0.5 cm
\begin{tabular}{cc}
\hskip -2 cm
\psfig{file=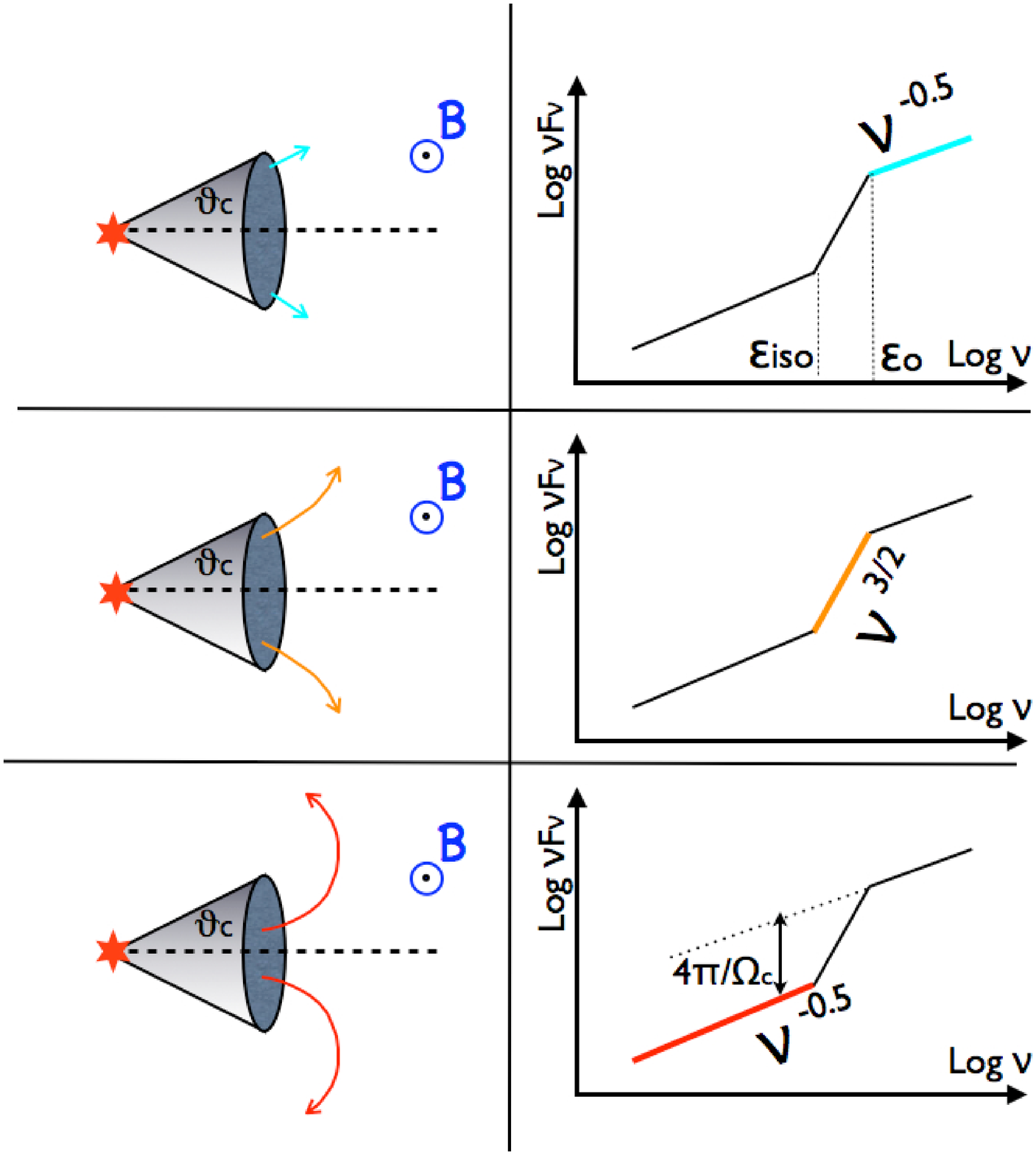,width=8cm,height=7.cm}
&\psfig{file=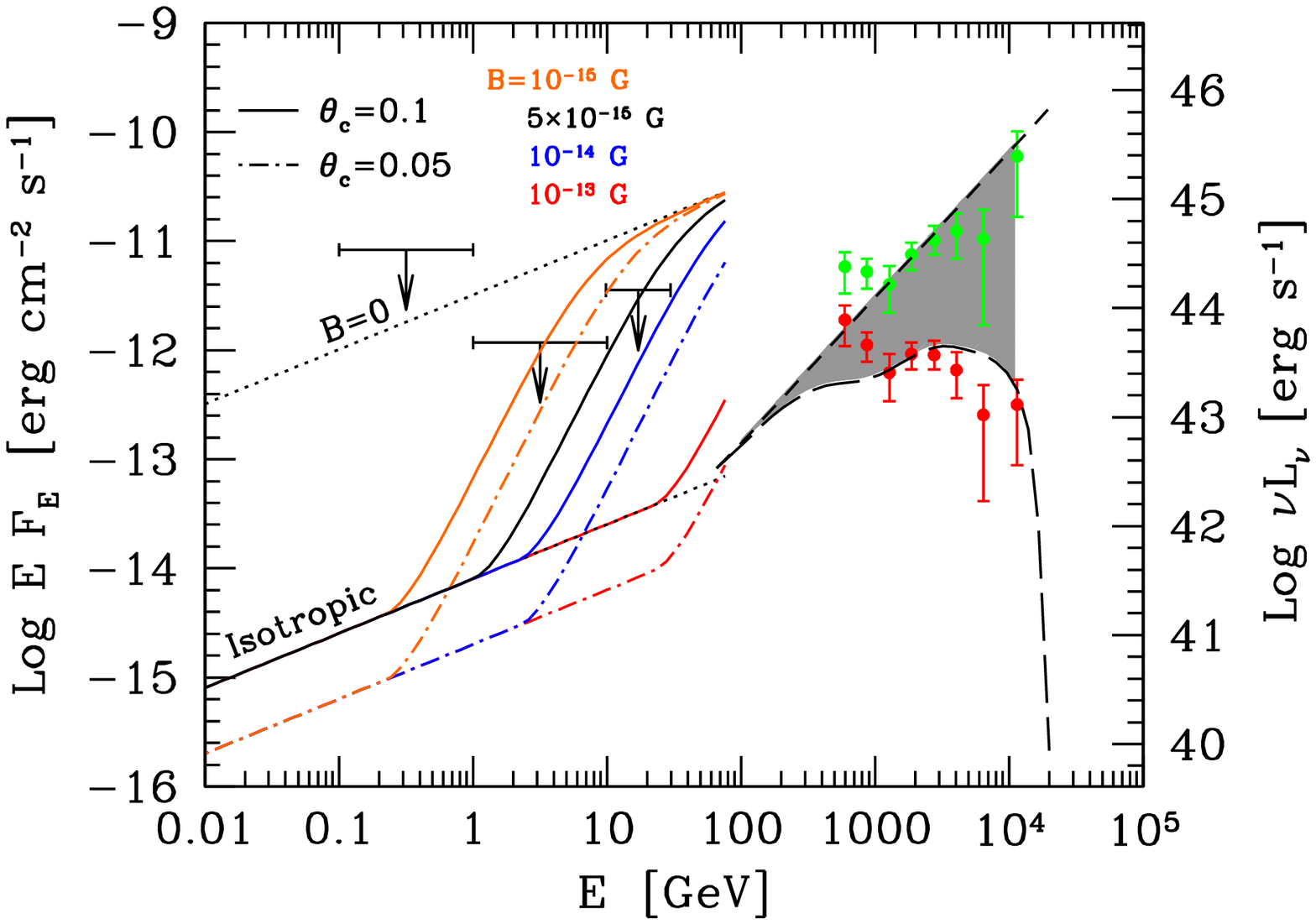,width=7cm,height=7.5cm}\\
\end{tabular}
\caption{
{\bf Left:}
Cartoon of the reprocessing of the absorbed TeV radiation.
TeV photons are emitted within a cone with semiÐaperture $\theta_{\rm c}$.
They are converted into pairs at a typical distance of hundreds of Mpc. 
Pairs cool rapidly through inverse Compton scattering on the CMB. 
If $B=0$, pairs do not gyrate, their emission is entirely contained in 
the original emission cone.
In this case the observed reprocessed radiation has the
same luminosity of the absorbed one.
If $B$ is non--zero, pairs with very high energy (top panel) cool before changing 
their direction (arrows) 
and thus their reprocessed emission is beamed within the same angle $\theta_{\rm c}$. 
Pairs of lower energies start to gyrate while cooling, and then emit within larger 
angles; thus their observed luminosity is smaller then the original one.
The resulting overall spectrum can be approximated by 
three power laws: i) electrons emitting within the original cone 
(top), ii) within a cone larger than the original (middle) and iii) almost isotropically (bottom).
{\bf Right:}
High energy SED of 1ES 0229+200.
Lower high energy points are the observed H.E.S.S. spectrum, (Aharonian et al. 2007) 
higher points are de--absorbed.
Lines are what predicted for different cosmic magnetic field values, as labelled,
and for the initial cone angle $\theta_{\rm c}$= 0.1 rad and 0.05 rad.
Black upper limits are from {\it Fermi}/LAT. From Tavecchio et al. 2010).
}
\label{cartoon}
\end{figure}

If the cosmic magnetic field is vanishingly small, the pairs will
emit their radiation within the original emitting cone of the blazar.
Therefore all the reprocessed radiation is contained in the
same cone of the original TeV radiation.
This implies that the flux is conserved: all the radiation
absorbed in the TeV energy range must come out at lower
frequencies.
If, instead, there is a non--negligible $B$--field, then
the pairs would gyrate along the field lines while they cool,
and the reprocessed radiation would be spread out in a cone
larger than the original blazar's one.
The received flux of the reprocessed radiation would then be lower.
The left panels of Fig. \ref{cartoon} sketch these cases.
Note that, in general, the high frequency reprocessed
radiation, corresponding to high energy particles and/or
the early phases of the particle cooling, is emitted 
in a short time, during which the particles do not gyrate much.
As the particle energy decreases, the cooling timescale gets longer,
the particle can gyrate more, and the reprocessed flux is spread 
within larger cones.
When the gyration is complete, the reprocessed radiation becomes isotropic.
We have therefore three regimes, corresponding to three different
spectral slopes of the reprocessed flux, as shown in Fig. \ref{cartoon}
(taken from Ref. \refcite{tavecchio2010}).
The right panel of the same figure illustrates an application 
to the TeV blazar 1ES 0229+200 ($z=0.14$), a particularly interesting
case because it has been observed up to $\sim10$ TeV.
The grey shaded region depicts the absorbed flux, and the different 
lines the expected level of the reprocessed flux for different cosmic 
$B$--fields.
The upper limits of the {\it Fermi} data limit the $B$--field to be {\it greater}
than $\sim 10^{-15}$ G. 
As pionereed by Ref. \refcite{neronov2010} and \refcite{tavecchio2010},
this technique is very promising, even if there are many complications
that one has to account for with respect to the heuristic explanation 
given above (see e.g. Ref. \refcite{dermer2011}).

\begin{figure} 
\centerline{\psfig{file=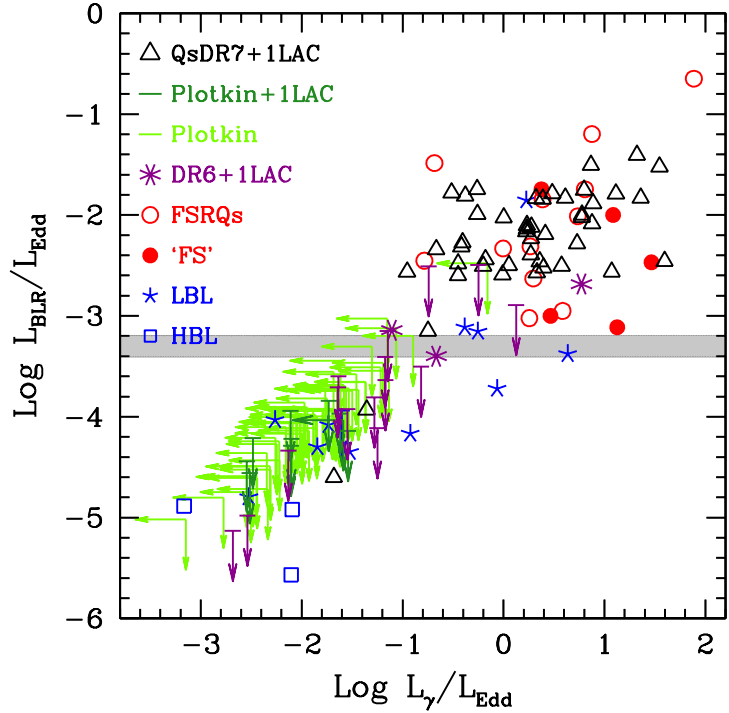,width=11cm}}
\vskip -0.5cm
\caption{
The luminosity of the BLR of blazars, in Eddington units,
as a function of their $\gamma$--ray luminosity in the {\it Fermi}/LAT energy band,
in Eddington units.
A strong (linear) correlation is present.
The grey horizontal stripe corresponds to $L_{\rm BLR}/L_{\rm Edd}=5\times 10^{-4}$,
and seems to divide (most of) BL Lacs from FSRQs. 
The upper limits on $L_{\rm BLR}/L_{\rm Edd}$ come from a sample of BL Lac objects present in the SDSS
for which Plotkin et al. (2011) could calculate the black hole mass through
the $M$--$\sigma$ relation.
They are all at $z<0.4$ and show absorption lines. From Sbarrato et al. (2011).
}
\label{tullia}
\end{figure}

\section{The FSRQ--BL Lac ``divide"}

The ``classic" and traditional way to classify blazars is through
the equivalent width (EW) of their broad emission lines.
FSRQs have EW larger than 5\AA\ (in the rest frame), while BL Lacs have lines
with EW
smaller than this value, or no lines at all (see e.g. Ref. \refcite{urry1995}).
This criterion is simple and of immediate practical use,
and it can be though of a sort of measure of the thermal to non--thermal ratio.
On the other hand, we know that the non--thermal jet continuum
is highly variable, and that most of the emission is at high
energies, not in the optical, and therefore the EW of optical
lines do not measure the ratio of the {\it bolometric} thermal
and non--thermal luminosities.

There are blazars classified as BL Lacs that have lines of small EW,
but whose luminosities is even larger than the one found in FSRQs.
Examples are PKS 0208--512\cite{scarpa1997}, 
AO 0235+164\cite{raiteri2007}, PKS 0426--380\cite{sbarufatti2005},
PKS 0537--441\cite{pian2002}.
Furthermore, the high amplitude optical variability makes some FSRQs
appear as BL Lacs when their non--thermal continuum is in the high state, and conversely, 
BL Lacs in a particular faint state can show broad emission lines that, albeit weak,
can have EW greater than 5 \AA\ (as BL Lac itself\cite{vermeulen1995}).

There is the need of a more physically based classification scheme, and we\cite{ghisellini2011} 
have proposed  a distinction based on the luminosity of the broad 
emission lines measured in Eddington units: $L_{\rm BLR}/L_{\rm Edd}$.
Normalizing to the Eddington luminosity ensures the appropriate comparison
among objects of different black hole masses.
We proposed that when $L_{\rm BLR}/L_{\rm Edd} \, \gsim \,5\times 10^{-4}$ 
the objects are FSRQs, and are BL Lacs below this value. 
The original sample of Ref. \refcite{ghisellini2011} was enlarged by 
Ref. \refcite{sbarrato2011}, confirming the earlier result.
Fig. \ref{tullia} shows the result of the latter study.
Note the large number of upper limits both in $L_{\rm BLR}/L_{\rm Edd}$ and
in $L_\gamma/L_{\rm Edd}$. 
These refer to BL Lac objects selected by Ref. \refcite{plotkin2011}
to be rather nearby ($z<0.4$), with signs of stellar absorption by stars of the host galaxy
in their SDSS spectra. 
They allowed Ref. \refcite{plotkin2011} to measure the black hole mass through the $M$--$\sigma$ relation.
Note that they all lie in the ``right" quadrant of Fig. \ref{tullia}, i.e. 
$L_{\rm BLR}/L_{\rm Edd}<5\times 10^{-4}$.
Fig. \ref{tullia} shows a good correlation between $L_{\rm BLR}$ and the $\gamma$--ray luminosity
in the {\it Fermi} energy range (the correlation remains essentially the same also when using
absolute values, not normalized to Eddington),
and a continuity between BL Lacs and FSRQs.

\subsection{ADAF?}

This poses a question: do the subdivision between BL Lacs and FSRQs reflects a fundamental
difference of only a different overall power (of both the jet and the accretion disk)?
We have proposed earlier (Ref. \refcite{ghisellini2009} for {\it Fermi} blazars;
Ref. \refcite{ghisellini2001} for FRI and FRII radio--galaxies) that one key 
and fundamental difference could be the accretion regime: FSRQs (with strong
emission lines) should have geometrically thin, optically thick accretion disks
accreting above $L_{\rm d}/L_{\rm Edd}=10^{-2}$, while BL Lacs could have
disk accreting below this value, and have Advection Dominated Accretion Flows 
(ADAF; e.g. Ref. \refcite{narayan1995}).
The found division at $L_{\rm BLR}/L_{\rm Edd}\sim 5\sim 10^{-4}$ is
roughly consistent with $L_{\rm d}/L_{\rm Edd}\sim 10^{-2}$,
especially if the covering factor of the BLR is $\sim$1/20.
This idea fits with the overall spectral properties of blazars:
strong and efficient disks ionize the broad line clouds, 
and the broad lines can be an efficient source of external seed photons
to be scattered by the jet to form the dominating high energy hump.
Weak and inefficient disks imply weak lines, fewer seed photons,
less cooling for the emitting electrons, and a weaker high energy hump,
produced mainly by the synchrotron self--Compton process,
but peaking at higher frequencies, because the emitting electrons,
suffering less cooling, can attain higher energies. 
Furthermore, as discussed more below, the power of the jet
likely correlates with the mass accretion rate: weaker disk have weaker jets.

On the other hand, the SED of ADAF is remarkably different from the
``multicolor" black--body of a standard\cite{ss1973} disk,
with a very reduced fraction of ionizing photons\cite{mahadevan1997}.
And yet we do observe broad emission lines even in rather extreme BL Lacs,
such as Mkn 501, Mkn 421 and PKS 2005--489 (see Ref. \refcite{ghisellini2011} and references therein), 
thought to have heavy black holes and very weak disk.
We do not have enough examples to draw any strong conclusion, but
note that the ADAF hypothesis, explaining the BL Lac/FSRQ divide, is not mandatory.
The transition to the ADAF regime could occur at much smaller values of 
$L_{\rm d}/L_{\rm Edd}$, as suggested by e.g. Ref. \refcite{sharma2007}.
One simple alternative is to assume that the location $R_{\rm diss}$ in the jet where
most of the radiation is produced is roughly the same in all blazars, once
measured in \sc\ radii (i.e.  $R_{\rm diss} \sim 10^3 R_{\rm S}$).
This is what comes out from fitting their SEDs\cite{ghisellini2010}.
The relation between the radius of the BLR and the ionizing luminosity
$R_{\rm BLR} \propto L_{\rm ion}^{1/2}$ (i.e. Ref. \refcite{kaspi2007}, \refcite{bentz2009})
means that a weak disk (even if not ADAF) implies a small $R_{\rm BLR}$.
If $R_{\rm BLR}$ becomes smaller than $R_{\rm diss}$, the external radiation, 
as seen in the jet comoving frame, becomes negligible.
In this case the division between BL Lacs and FSRQs is still dependent on $L_{\rm d}/L_{\rm Edd}$
but, in addition, there is also a dependence on the black hole mass,
making the expected divide more blurred (but still consistent with existing 
data\cite{sbarrato2011}).

\section{Jet and accretion power}

Since the jet radiation is beamed, it is not trivial to estimate the jet power from 
what we see.
In a pioneering study\cite{rawlings1991},
the average jet power was estimated from the extended radio emission, calculating
the minimum  (equipartition) energy of the radio lobes (or extended structures), and
then dividing by an estimate of the source lifetime.
Very interestingly, the jet power was found to linearly correlate with the luminosity
of the narrow emission lines (thought to be produced by photo--ionization by the
accretion disk), suggesting an approximate equality of the jet power and the
disk luminosity.
At large scales, the jet power can be measured in optical (by {\it HST}) and/or X--rays (by {\it Chandra}).
Another promising way to measure the jet power is to use the X--ray bubbles/voids seen in 
the intra--cluster medium hosting radio-AGNs, allowing to measure the $PdV$ work, the age, 
and then the jet power\cite{allen2006}\cdash\cite{balmaverde2008}.

However most results on the power of blazars' jet consider the zone of the jet where most
of the radiation is produced, assuming that this zone is a single one (a blob), characterized
by the same magnetic field, particle density and bulk motion.
Through modelling, we can derive how many leptons and magnetic field are required 
to account for the observed emission, what is the size, and the bulk Lorentz factor.
The jet power is in several forms, but all can be calculated through:
\begin{equation}
P_{\rm i} = \pi R^2\Gamma^2 \beta c U^\prime_{\rm i}
\end{equation}
where $R$ is the size of the emitting blob, $U^\prime_{\rm i}$ is
a comoving energy density.
The different forms of power are the magnetic one ($U^\prime=U_{\rm B}$),
the kinetic power of the emitting electrons 
($U^\prime=U_{\rm e}=n_{\rm e}\langle\gamma\rangle m_{\rm e}c^2$), the kinetic power of the 
cold protons ($U^\prime=U_{\rm p}=n_{\rm p} m_{\rm p}c^2$), and the power 
in the produced radiation [$U^\prime=U_{\rm rad} \sim L^\prime_{\rm bol} /(4\pi R^2 c)$].
To calculate the latter we need the observed bolometric luminosity and an estimate of the
bulk Lorentz factor:
\begin{equation}
P_{\rm r} = \pi R^2\Gamma^2 \beta c {L^\prime_{\rm bol} \over 4\pi R^2 c }
= {\Gamma^2 L^{\rm obs}_{\rm bol}\over 4 \delta^4}
\sim {L^{\rm obs}_{\rm bol}\over 4 \Gamma^2}
\end{equation}
where the last equality assumes that the Doppler factor $\delta=\Gamma$.
This is a {\it lower limit} to the total jet power $P_{\rm jet}$,
since the jet would stop if all its (kinetic+Poynting) power is used
to produce the radiation we see (even if in extreme TeV BL Lacs this can be exactly what
happens: this would explain why $\Gamma$ is very large in the radiative region,
but small on the larger VLBI scale as indicated by 
mildly superluminal, or even sub--luminal apparent speeds; 
see Ref. \refcite{ghisellini2005}, 
\refcite{markos2003}).

\begin{figure}[h]
\begin{tabular}{cc}
\hskip -0.5 cm
\psfig{figure=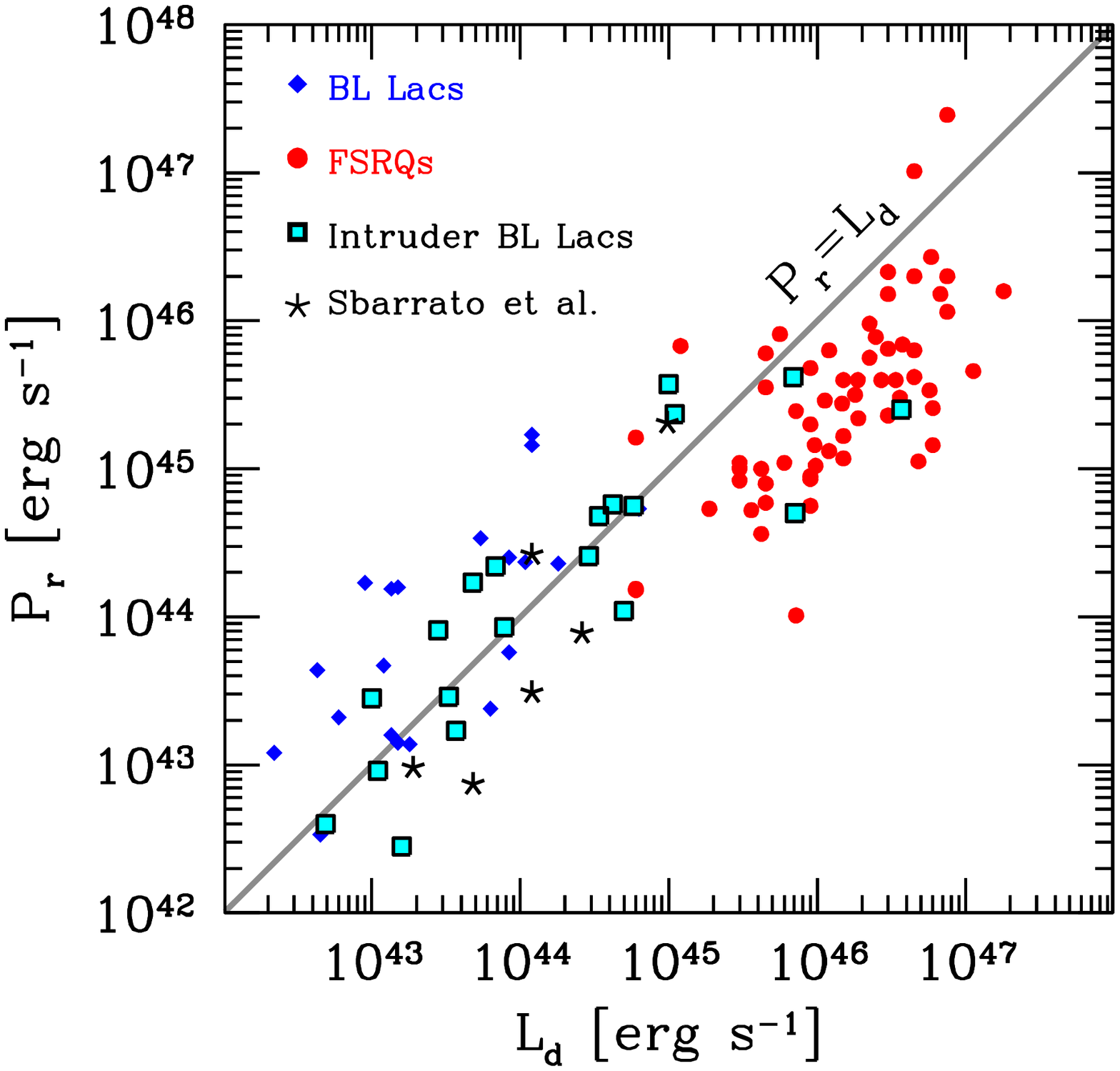, width=6.5cm} 
&\psfig{figure=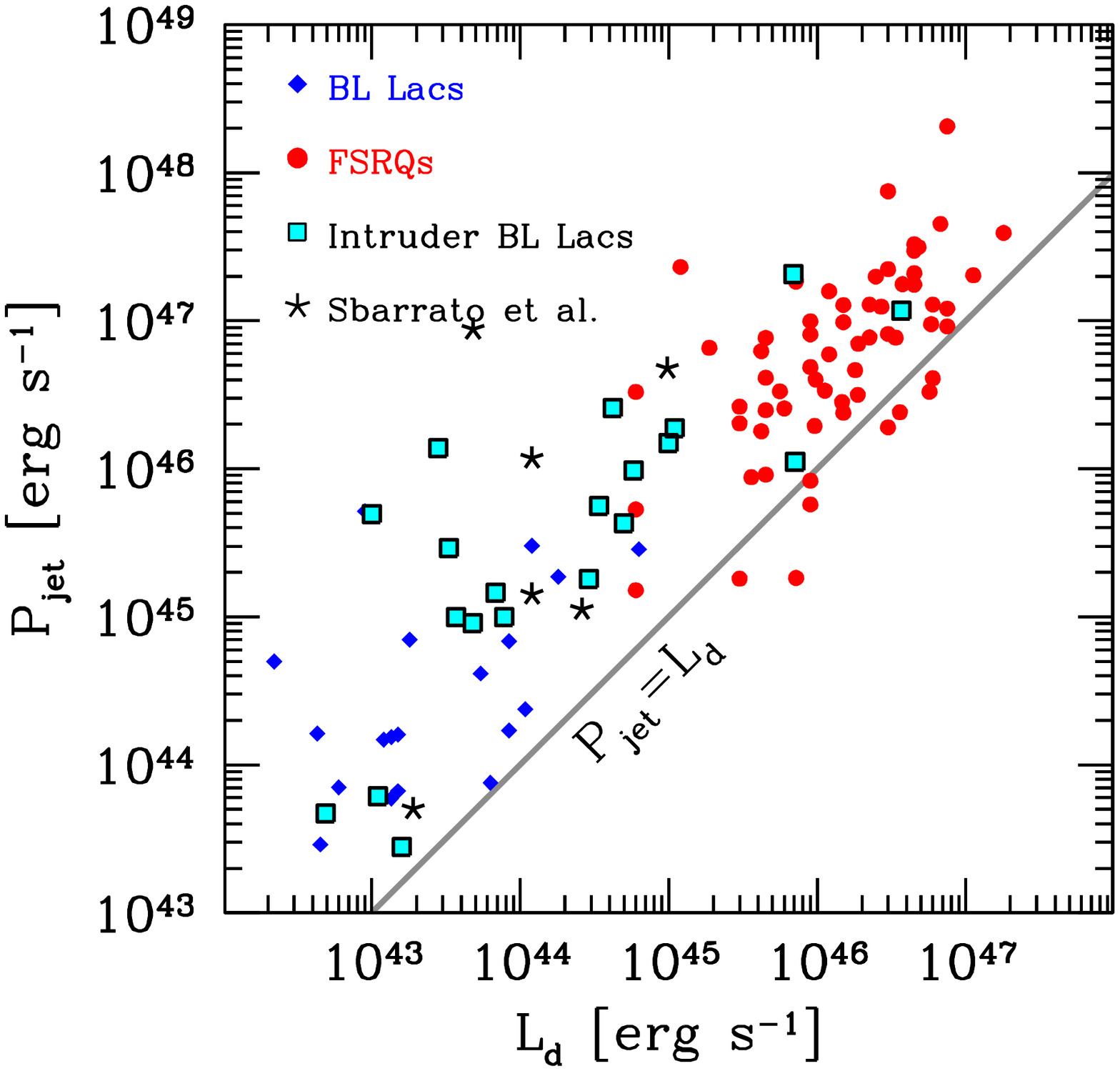, width=6.5cm} \\
\end{tabular}
\vskip -0.3cm
\caption{
The power spent by the jet to produce the radiation we see, $P_{\rm r}$ (left)
and the total jet power $P_{\rm jet}$ (right) as a function of the accretion 
disk luminosity $L_{\rm d}$, calculated through the observed broad emission lines 
(assuming $L_{\rm d} =10 L_{\rm BLR}$).
$P_{\rm r}$ can be considered as a robust lower limit to $P_{\rm jet}$, which is here
calculated assuming one cold proton per emitting electron.
Grey solid lines indicate equality.
}
\label{ld}
\end{figure} 

Fig. \ref{ld} shows $P_{\rm r}$ (left panel) and $P_{\rm jet}$ (right) as a function
of the accretion disk luminosity $L_{\rm d}$ for a sample of {\it Fermi} blazars.
It contains both FSRQs and BL Lacs.
For FSRQs the accretion disk component can be often seen directly
in the optical--UV range; for BL Lacs it is deduced mainly from the luminosity of the (weak)
broad emission lines.
Fig. \ref{ld} shows that: i) the jet power is large; ii) it nicely correlates with
$L_{\rm d}$; if one takes one proton per emitting electron (see Ref. \refcite{celotti2008}, 
\refcite{sikora2000} for a discussion of this issue), then $P_{\rm jet}>L_{\rm d}$,
on average, suggesting that
\begin{equation}
P_{\rm jet} \sim \dot M c^2
\end{equation}
for all values of $\dot M$. Sometimes, it can be even larger (it surely is during exceptional
flares, as in 3C 454.3\cite{bonnoli2010}\cdash\cite{abdo2011}).
The fact that $P_{\rm jet}$ correlates with $L_{\rm d}$, but that it can be larger than that, is an 
apparent paradox: the correlation would point to accretion as the prime mover of jets, but then
how can $P_{\rm jet}$ be larger than $L_{\rm d}$?
This paradox can be solved by noting that, to extract the black hole rotational energy,
we need a strong magnetic field, that has to be produced/sustained in the 
inner parts of the accretion disk.
We then need $B_0^2 \sim 8\pi \rho_0 c^2$ in the innermost regions of the
disk to make the Blandford--Znajek\cite{blandford1977} process efficient.
This links the jet and the accretion powers.
This is also borne out by recent numerical simulations (e.g. Ref. \refcite{tchekhovskoy2011}), showing 
that, occasionally, $P_{\rm jet}$ can be even larger than $\dot M c^2$.

\begin{table}[ph]
\tbl{Summary of the average physical properties of {\it Fermi} bright blazars. 
Values from Ghisellini et al. (2010).
$R_{\rm diss}/R_{\rm s}$ is the distance of the dissipation region from the central 
black hole of mass $M$ (in \sc\ radii). 
$\Gamma$ is the bulk Lorentz factor. 
$B$ is the magnetic field in the emitting region, as measured in the comoving frame.
$L_{\rm d}$ is the luminosity of the accretion disk.
$P^\prime$ is the power injected in the form of relativistic electrons, measured in the comoving frame.
$P_{\rm jet}$ is the total power of the jet, assuming one proton per emitting electron.
}
{\begin{tabular}{@{}lllllllll@{}} 
\hline
\hline
~ &$\log M$    &$R_{\rm diss}/R_{\rm S}$ &$\Gamma$ &$B$ &$\log L_{\rm d}$ &$L_{\rm d}/L_{\rm Edd}$ &$\log P^\prime$  &$\log P_{\rm jet}$ \\
  &[$M_\odot$] &                       &         &G   &[erg/s]   &                                    &[erg/s]   &[erg/s]              \\
\hline
BL Lacs &8--9    &300-1000  &10--20 &0.1--2 &42--44   &$<10^{-2}$ &41--43   &43.5--45 \\
FSRQs   &8--9.5  &300--3000 &10--16 &1--10  &44--46.5 &$>10^{-2}$ &42.5--44 &45--48 \\
\hline
\hline
\end{tabular} 
\label{tab}}
\end{table}

\begin{figure} 
\centerline{\psfig{file=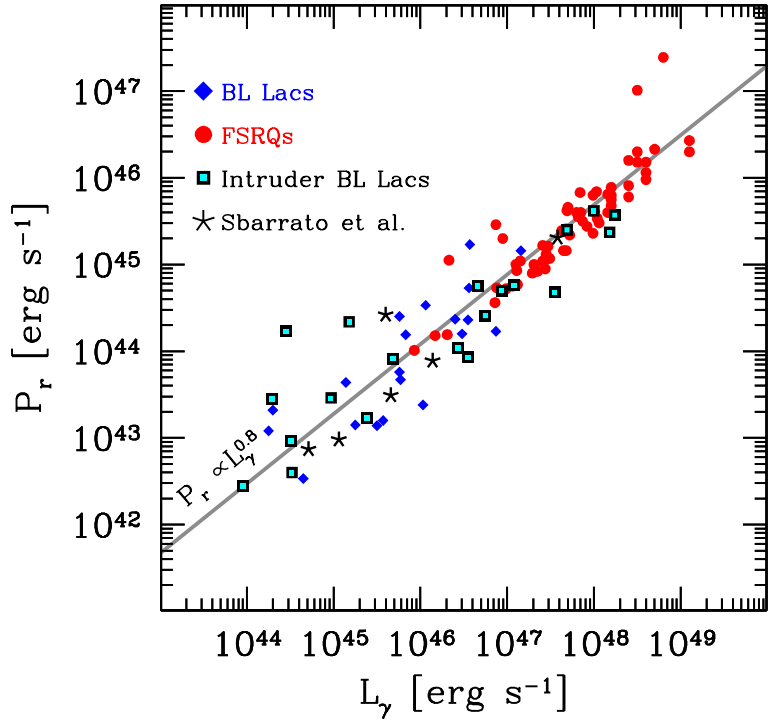,width=11cm}}
\vskip -0.5cm
\caption{
The power $P_{\rm r}$ that the jet spends to produce the radiation we see as a function of 
the $\gamma$--ray luminosity in the {\it Fermi}/LAT energy band (K--corrected).
The bisector of the least square best fit yields $P_{\rm r} \propto L_\gamma^{0.8}$
(and $L_\gamma \propto P_{\rm r}^{1.27}$).
The non linearity is mainly due to the fact that for low power BL Lacs the $\gamma$--ray
luminosity (as measured in the {\it Fermi}/LAT energy band underestimate the
bolometric one.
}
\label{lgpr}
\end{figure}

\section{Conclusions}

Tab. \ref{tab} summarizes the average physical properties of
BL Lacs and FSRQs, derived from fitting their overall SED 
(including the $\gamma$--rays).
The main difference is their overall power (both of their jets and their disks),
inducing a difference in the value of the magnetic field in the
jet emitting region.
Since the black hole masses are similar, the difference of the accretion power
means a difference in the $L_{\rm d}/L_{\rm Edd}$ ratio, and a well defined divide.
The division between BL Lacs and FSRQs occurs at a given luminosity of the broad 
emission lines measured in Eddington units: $L_{\rm BLR}/L_{\rm Edd}=5\times 10^{-4}$,
which corresponds to $L_{\rm d}/L_{\rm Edd}=10^{-2}$ if the covering factor of the BLR
is $\sim 1/20$.
The changing regime of accretion (from radiatively inefficient to efficient)
can be the explanation of the different ``look" of BL Lacs and FSRQs.
Alternatively\cite{sbarrato2011}, the disk could remain radiatively efficient even for 
small values of $L_{\rm d}/L_{\rm Edd}$, and in this case the BL Lac/FSRQ ``divide"
could be due to the emitting region being outside (BL Lac) or inside (FSRQs)
the Broad Line Region, whose size scales as $L_{\rm d}^{1/2}$.

A useful lower limit to the jet power is $P_{\rm r}$, the
power spent to produce the radiation we see. 
It is a robust estimate because it depends on the 
observed bolometric luminosity and the bulk Lorentz factor, so it is nearly
model--independent. 
Since the $\gamma$--ray luminosity is often the dominant contribution
to the bolometric one, instead of $L_{\rm bol}$
one can use $L_\gamma$, and the following relation:
\begin{equation}
P_{\rm r} = 1.3\times 10^7 L_\gamma^{0.8} \quad {\rm erg \, s^{-1}}
\end{equation}
where $L_\gamma$ is the K--corrected 0.1--10 GeV luminosity.
This relation is shown in Fig. \ref{lgpr} together with the 
observed points.
It can be used to quickly estimate a lower limit to the jet power,
bearing in mind that i) the $\gamma$--ray luminosity can vary by 
even 2 orders of magnitude  and ii) that the real jet power $P_{\rm jet}$
can much larger ($\sim$ 2 orders of magnitude) than $P_{\rm r}$, 
if charge neutrality is due to cold protons.

Blazars (and then their misaligned counterparts) have jets
for a large range of accretion disk luminosities,
in Eddington units.
They range from $\sim$1 (close to Eddington) to $\sim 10^{-4}$.
The SED of blazars is often dominated by the high energy hump,
with the synchrotron luminosity being a minor part in FSRQs.
This limits the value of the magnetic field.
Indeed, it turns out that the Poynting flux in the emitting
region is less than $P_{\rm r}$.
This means that in FSRQs the jet power is mainly in the bulk
motion of cold protons.

On the other hand, the only option to accelerate jet in blazars
is through the Poynting flux:
contrary to Gamma Ray Bursts, the jets of blazars are initially optically thin.
The scattering optical depth $\tau_{\rm T}$ is less than unity. 
Even if  e$^\pm$ pairs were copiously produced, 
they would quickly cool and annihilate, making $\tau_{\rm T}\lsim$1
at the very start.
This means that any internal radiation cannot be transformed into bulk
motion (this can instead occur in Gamma Ray Bursts).
Internal energy in the form of very hot (i.e. $\gamma\gsim 15$)
matter would again very quickly cool in the dense radiation field 
produced by the nearby accretion disk.
Again, we cannot use this internal energy to accelerate blazars' jets.
The only remaining option is then a strong initial Poynting flux, that
must be dominant at the start, and yet it must become a minor component in 
the emitting region, $\sim 10^3$ \sc\ radii away.

\section*{Acknowledgments}

I thank G. Bonnoli, A. Celotti, L. Foschini, G. Ghirlanda, L. Maraschi, 
T. Sbarrato and F. Tavecchio for many fruitful discussions, and
J. Krolik for interesting comments.
I thank the PRIN-INAF 2009 grant for funding.


\end{document}